\documentclass[aps,showpacs,superscriptaddress,nofootinbib,amsmath,amssymb]{revtex4}
 \usepackage{graphicx}
\usepackage{dcolumn}
\usepackage{bm}
\usepackage{lipsum}
\usepackage{adjustbox}
\usepackage{multirow}
\usepackage{amsmath}
\usepackage{amssymb}
\usepackage{enumerate}

\graphicspath{./fig}
\begin{document}
\title{Nuclear medium effects in lepton-nucleus DIS in the region of $x \gtrsim 1$}
\author{M. Sajjad Athar}
\author{S. K. Singh}
\author{F. Zaidi\footnote{Corresponding author: zaidi.physics@gmail.com}}
\affiliation{Department of Physics, Aligarh Muslim University, Aligarh - 202002, India}
\begin{abstract}
The nuclear medium effects in the nuclear structure functions and differential cross sections in the deep inelastic scattering (DIS) of charged lepton and neutrino from 
nuclear targets are studied in the region of large $x$ including $x\ge 1$. The nuclear medium effects due to the Fermi motion and the binding energy of nucleons and the nucleon correlations are 
included using nucleon spectral function calculated in a microscopic field theoretical model. The numerical results for the nuclear structure functions 
and the cross sections are obtained using the nucleon structure function evaluated at the next-to-next-to-leading order (NNLO) with
the Martin-Motylinski-Harland Lang-Thorne (MMHT) parameterization of the nucleonic parton distribution functions (PDFs) and are compared with the available experimental data on electron 
scattering from the Jefferson Lab (JLab) and SLAC Nuclear Physics Facility (NPAS). In the case of neutrino scattering the results are relevant for understanding the DIS contributions to
the recent inclusive cross sections measured by the 
Main Injector Neutrino ExpeRiment to study v-A interactions (MINERvA) as well as theoretical predictions are made for Deep Underground Neutrino Experiment (DUNE). The importance of isoscalarity corrections in heavier nuclear targets as well as the effect of 
the kinematic cut on the CM energy $W$ in defining the DIS region have also been discussed.
\end{abstract}
\pacs{13.15.+g,13.60.Hb,21.65.+f,24.10.-i}
\maketitle

\section{Introduction}
The experimental evidence of the nuclear medium effects in the deep inelastic scattering (DIS) of the charged leptons from the heavier nuclear targets was first reported by the European Muon Collaboration (EMC)
when it measured the cross sections for $\mu^--^{56}Fe$ and $\mu^--^{2}D$ scattering processes and found that the ratio of the cross sections per nucleon in $^{56}Fe$ and $^{2}D$ is not unity~\cite{Aubert:1983xm}.
As the DIS cross sections are generally expressed in terms of the nucleon structure functions, the EMC observation implied that the structure functions for a nucleon bound inside a nucleus are different from the structure functions of a free nucleon. This was surprising as the underlying degrees of freedom participating in the DIS process are quarks and gluons. This effect is 
famously known as the EMC effect.
Later more experiments confirmed this EMC observation for $\Big(\frac{A}{A'}\Big)\Big(\frac{\sigma_{A'}}{\sigma_{A}}\Big)$, where $A$ and $A'$ are the nucleon number for the different 
nuclear targets~\cite{Gomez:1993ri, NewMuon:1995cua, Arneodo:1995cs, Ackerstaff:1999ac, Benvenuti:1987az, Bari:1985ga, Seely:2009gt}.
The deviation of the ratio $\Big(\frac{A}{A'}\Big)\Big(\frac{\sigma_{A'}}{\sigma_{A}}\Big)$ 
from unity underlines the importance of the nuclear medium effects through the structure of the EMC effect and is categorized in four broad categories depending upon the different
regions of $x$, for example, the shadowing and antishadowing effects in the region of $0<x<0.1$ and $0.1<x<0.2$, the EMC effect in the region of $0.2<x<0.7$ and the 
Fermi motion effect in the region of $x\ge 0.7$. For the free nucleon target, the Bjorken variable $x$ lies in the region of $0\le x\le 1$ while for a nuclear target it can vary 
from 0 to $A$, i.e. $0\le x \le A$. To explore the region of $x>1$ lepton induced deep inelastic scattering from a nuclear target is the conventional method and the quarks carrying momentum fraction greater than the
momentum of nucleon at rest are referred as the ``superfast quarks'' in the literature. These superfast quarks may be described via understanding the different properties of quantum chromodynamics (QCD) like the behavior of nuclear
forces at very short distances and the nuclear medium modifications of parton distributions in nuclei. Several nuclear PDFs parameterizations are available in the literature and continuously being
updated~\cite{Segarra:2020gtj, Freese:2015ebu, Eskola:2021nhw, AbdulKhalek:2018vqb, Khanpour:2016pph, Hirai:2007sx, deFlorian:2011fp}.
In the region of $x$ larger than 1, i.e., $x\ge 1$ (inaccessible for the free nucleons), some experimental efforts have been made to study the charged lepton-nucleus inclusive scattering
processes~\cite{Arrington:2021vuu, Arrington:1995hs,Arrington:2001ni,CLAS:2005pgc, CLAS:2010nnk, Filippone:1992iz, BCDMS:1994ala},
while in the weak sector, not many studies have been made to explore the region of $x\ge 1$~\cite{MINERvA:2014rdw, CCFR:1999kbf}. 

The neutrino-nucleus scattering experiment MINERvA at the Fermilab~\cite{MINERvA:2014rdw, Mousseau:2016snl} 
has been performed to make EMC kind of measurements using medium and heavy nuclear targets like hydrocarbon, water, iron and lead in a wide range of $x$ and $Q^2$. 
This experiment is not only giving the information on the hadron dynamics in the nuclear medium for weak interaction induced processes in the wide region of $x$ and $Q^2$ but it would also be helpful in
understanding the nuclear model dependence of $\nu_l(\bar\nu_l)-A$ scattering cross section. This is important for interpretation of the neutrino
oscillation experiments being done using nuclear targets. A better understanding of 
scattering cross section is required to reduce the systematics which has presently 25-30\% contribution due to the lack in the understanding of neutrino-nucleus scattering cross sections. In the first results from MINERvA collaboration,
Tice et al.~\cite{MINERvA:2014rdw} have reported the experimental results for the ratios of differential cross section, i.e, $\Big(\frac{d\sigma_A/dx}{d\sigma_{CH}/dx}\Big);~(A=^{12}C,~^{56}Fe,~^{208}Pb)$
for the $\nu_\mu-A$ inclusive scattering processes using the low energy neutrino beam (peaks around neutrino energy $E_{\nu_l}=3$ GeV) in the energy region of $2\le E_{\nu_l}\le 20$ GeV in a wide range of $x$, i.e.
$x\le 1.5$. They observed that theoretical models proposed to include the nuclear medium effects are not able to explain the MINERvA's experimental data~\cite{MINERvA:2014rdw}. Hence, the high statistics experimental measurements as well as better theoretical understanding of nuclear medium effects 
are required. Presently, the experimental analysis of the (anti)neutrino induced inclusive scattering data using intermediate energy (anti)neutrino beam in the energy
region of $5-50$ GeV peaking at the (anti)neutrino energy $E_{\nu_l}=6$ GeV is under process. This would provide new data in a wide range of $x$, including the higher region of $x\gtrsim 1$~\cite{DanR}.
Moreover, the current and future experiments at the Fermilab with the short-baseline and long-baseline neutrino beams like ICARUS~\cite{Tortorici:2019mwg, Machado:2019oxb}, 
SBND~\cite{MicroBooNE:2015bmn, SBND:2020scp}, MicroBooNE~\cite{MicroBooNE:2021cue} and DUNE~\cite{Abi:2018dnh, Abi:2020mwi, Abi:2020qib} are also aiming the measurements of neutrino-nucleus
scattering cross sections, specifically, using the liquid argon as a nuclear target. The liquid argon scintillators are being used due to their capability of excellent 
 neutrino flavor identification and neutrino energy reconstruction. ICARUS~\cite{Tortorici:2019mwg, Machado:2019oxb}, SBND~\cite{MicroBooNE:2015bmn, SBND:2020scp} and MicroBooNE~\cite{MicroBooNE:2021cue} experiments are focused to explore the 
energy region of a few GeV, however, DUNE experiment~\cite{Abi:2018dnh, Abi:2020mwi, Abi:2020qib} has a wide energy spectrum which spans up to tens of GeV. Earlier ArgoNeuT
experiment~\cite{ArgoNeuT:2011bms} was performed
at the Fermilab using the NuMI neutrino beamline which first reported the results of the inclusive cross sections for mean neutrino energy $<E_{\nu_l}>=4.3$ GeV in 
the charged current $\nu_\mu-^{40}Ar$ scattering process. Later, the 
measurements were also made for the $\bar\nu_\mu$ induced inclusive scattering off argon with a mean antineutrino energy of 3.6 GeV~\cite{ArgoNeuT:2014rlj}.

In the energy region of a few GeV, the contribution to the neutrino-nucleus cross section 
comes from the quasielastic, resonance production and deep inelastic scattering processes. It is not easy to exactly define the kinematic regions corresponding to these processes but one can classify them depending
upon the dominance of a particular process. For example, in the low energy region $E_{\nu_{l}}< 1$ GeV it is the quasielastic scattering which gives dominant contribution
to the cross section and in the energy region of $1\le E_{\nu_{l}}\le 3$ GeV the inelastic scattering processes dominate while in the energy region of $E_{\nu_{l}}>3$ GeV the DIS dominates. 
However, there is significant contributions of the higher resonance production in the region of $E_{\nu_{l}}\ge 3$ GeV, while the contribution of the DIS in the energy region of $E_{\nu_{l}}\le 3$ GeV is also
not small. In view of this the kinematic region around $E_{\nu_{l}}\approx 3$ GeV is designated as the transition region of resonance production and DIS.
The sharp kinematic boundaries defining the transition regions are not defined uniquely. Therefore, in the literature to define the DIS region, the kinematic constrain on $W$ has varied from 1.4 to 2.0 GeV and it has been extrapolated to lower values of $Q^2<1$ GeV$^2$~\cite{Kretzer:2002fr, Paschos:2001np, Jeong:2010nt, Hagiwara:2003di}.
Due to the ambiguity in the definitions of the transition region, there exists an uncertainty in calculating the total cross sections while summing over the cross section contributions from the 
resonance production and DIS processes. Hence, it is important to properly define the kinematic boundaries for the transition region. In MINERvA's experimental analysis the region 
of $Q^2>1$ GeV$^2$ and $W>2$ GeV is considered to be the region of pure or true DIS~\cite{Mousseau:2016snl}. The same kinematic constrains to define the region of pure DIS have been used 
in the neutrino event generators such as NEUT~\cite{Hayato:2009zz} and GENIE Monte Carlo~\cite{Andreopoulos:2009rq} which is widely used by the neutrino 
physics community. Moreover, the neutrino-nucleus interactions in the shallow and deep inelastic scattering regions have also been discussed in the NuSTEC workshop 
held at L$'$Aquila in 2018~\cite{NuSTEC:2019lqd} as well as in the Snowmass conference held in 2021~\cite{Alvarez-Ruso:2020ezu}. The understanding of neutrino physics in these kinematic regions
is important in order to interpret the experimental results from the current and future oscillation experiments using accelerator and atmospheric neutrinos. In the review article by 
Athar et al.~\cite{SajjadAthar:2020nvy} and also in Ref.~\cite{Zaidi:2021iam}, the transition region and 
corresponding kinematical constrains are discussed in detail.

\begin{figure}[h]
 \includegraphics[height=6.5 cm, width=7.5 cm]{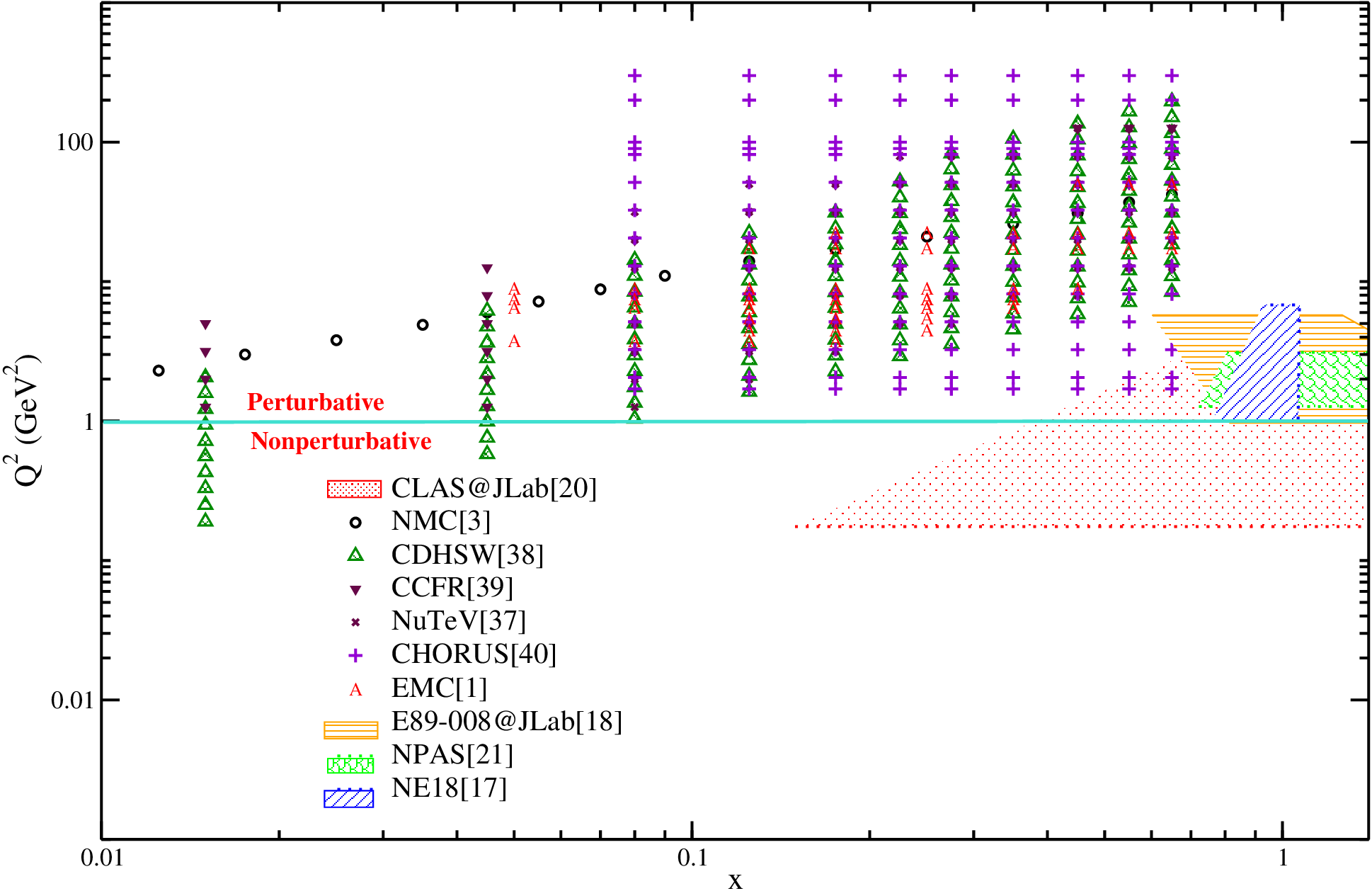}
 \includegraphics[height=6.5 cm, width=7.5 cm]{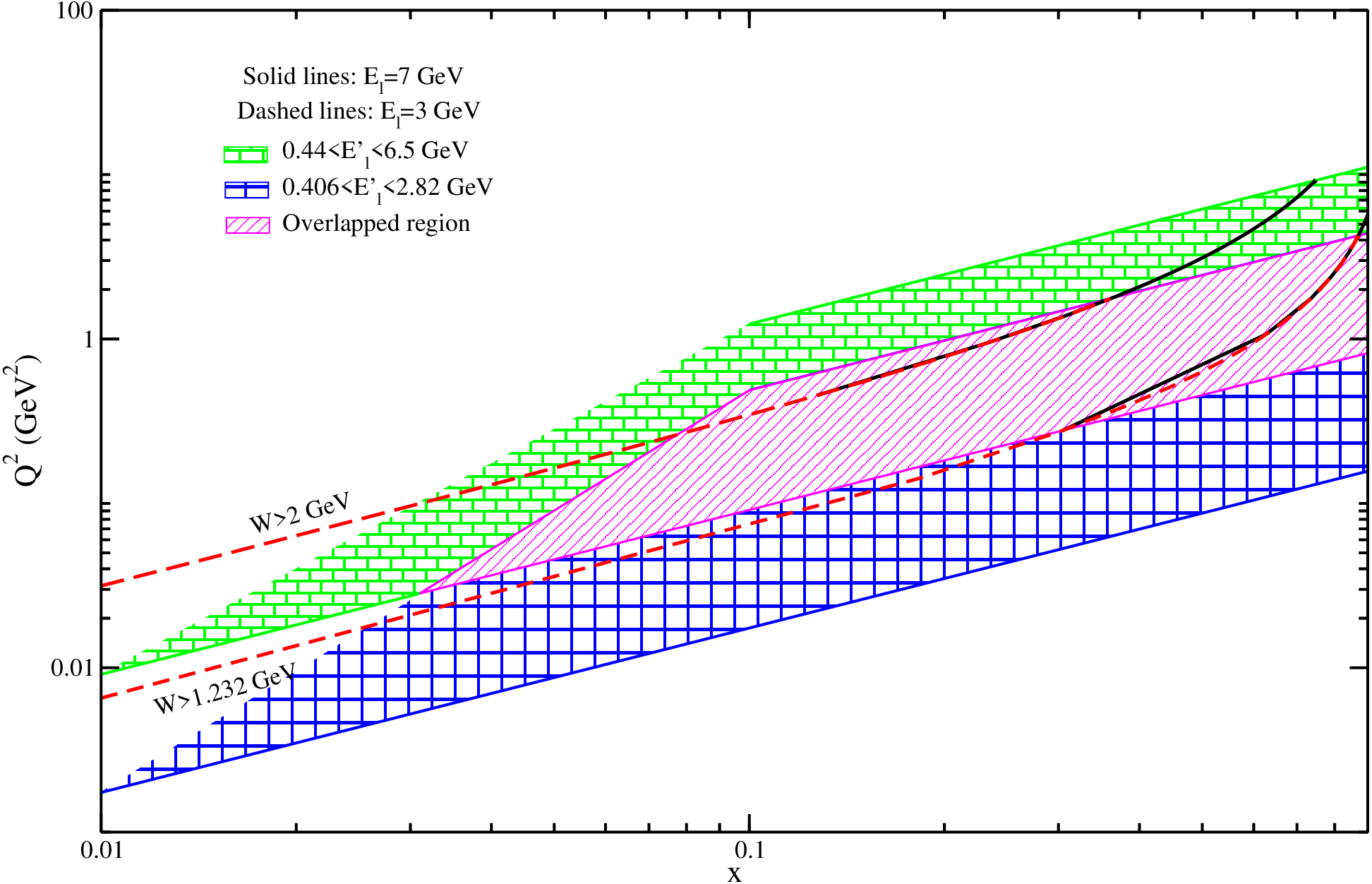}
\caption{{\bf Left panel:} Kinematic region in $x-Q^2$ plane, where most of the experimental data are available from charged 
lepton-nucleus~\cite{Aubert:1983xm, NewMuon:1995cua, CLAS:2010nnk, Arrington:1995hs, Arrington:2001ni, Filippone:1992iz}
and neutrino-nucleus~\cite{Tzanov:2005kr, Berge:1989hr, Oltman:1992pq, Onengut:2005kv} scattering off various 
nuclei like $^{12}C$, $^{56}Fe$ and $^{208}Pb$. The data from CLAS (red dotted pattern), NE18 (blue diagonal lines pattern), E89-008 (orange horizontal lines pattern) and NPAS (green down wave pattern) 
experiments are shown by the band. {\bf Right panel:} The two bands in the figure show the kinematic region of outgoing charged lepton energy ($E_l'$) 
in the range of $0.406<E_l'<2.82$ GeV (square pattern) and $0.44<E_l'<6.5$ GeV (brick pattern) corresponding to the incoming beam energies viz. $E_l=3$ GeV and $E_l=7$ GeV, respectively. The overlapped region of these energies 
is shown by the band filled with diagonal lines. The dashed (solid) lines show the $x-Q^2$ region for $E_l=3 (7)$ GeV when a cut on the CM energy viz. $W>1.232$ GeV and $W>2$ GeV is applied. The 
kinematic region of $x>1$ which is inaccessible for the free nucleon target (right panel) may be explored using a nuclear target (left panel) as shown in the present figure. Therefore, the
investigation of deep inelastic scattering in the kinematic region $x>1$ has significant importance in the understanding of nuclear medium effects. }
\label{xq2_fig}
\end{figure}

In Fig.~\ref{xq2_fig}, we have shown the kinematic region of $x$ and $Q^2$ covered by 
the neutrino-nucleus scattering experiments viz. NuTeV~\cite{Tzanov:2005kr}, CDHSW~\cite{Berge:1989hr}, CCFR~\cite{Oltman:1992pq}
and CHORUS~\cite{Onengut:2005kv} and some of the charged lepton-nucleus scattering experiments viz. EMC~\cite{Aubert:1983xm}, NMC~\cite{NewMuon:1995cua}, NE18~\cite{Arrington:1995hs}, E89-008 at JLab~\cite{Arrington:2001ni},
CLAS~\cite{CLAS:2010nnk}, NPAS~\cite{Filippone:1992iz}. One may notice that there is lack of 
experimental data in the region of $x>0.8$ for both the electromagnetic as well as weak interaction channels. 
Therefore, more experimental measurements are required to explore the nucleonic properties and the lepton production cross section in this kinematic region. In this figure, we have also shown
the corresponding kinematic regions in the $x-Q^2$ plane for the free nucleon target (right panel) at the two different values of incoming lepton energies (either neutrino or charged lepton)
viz. $E_l=3$ GeV and $E_l=7$ GeV with 
outgoing lepton energies $E_l'$ lying in the range of $0.406\le E_l'\le 2.82$ GeV and $0.44\le E_l'\le 6.5$ GeV, respectively. The effect of the center of 
mass energy cuts of $W>1.2$ GeV and $W>2$ GeV has also been shown on the allowed kinematic region in the $x-Q^2$ plane as these quantities are related as:
\begin{equation}
 x=\frac{Q^2}{2 M_N (E_l-E_l')},~~~Q^2=-q^2\ge0,~~~W^2=M_N^2+Q^2\left(\frac{1}{x}-1\right)
\end{equation}
with $M_N$ as the target nucleon mass. 

In the considered kinematic region of high $x$ and moderate $Q^2$ the nonperturbative effects like the target mass corrections (TMC) and higher twist corrections (HT) that involve 
the powers of $\left(\frac{1}{Q^2}\right)^n; ~n=1,2,3...,$ become important.
Therefore, nonperturbative effects are of considerable experimental interest to the oscillation experiments.
Furthermore, the higher order perturbative evolution of parton densities is also an important aspect in this kinematic region of $x$ and $Q^2$, where due to finite value of the 
strong coupling constant ($\alpha_s(Q^2)$), terms beyond the leading order (LO) can not be ignored.  
In this paper, we have performed the numerical calculations up to next-to-next-to-leading order (NNLO) 
by using the MMHT nucleonic PDFs parameterization~\cite{Harland-Lang:2014zoa}. In addition to that we have taken into account the TMC effect following the works 
of Schienbein et al.~\cite{Schienbein:2007gr}. 

Theoretical investigation of nuclear medium effects in the DIS region for $x$ beyond 0.8 is limited in the literature~\cite{Saito:1985ct, Frankfurt:1988nt, FernandezdeCordoba:1995pt}, especially
in the case of neutrino interactions with the nuclear targets. Bodek and Ritchie~\cite{Bodek:1981wr, Bodek:1980ar} have reported the effect of Fermi motion on the weak
nuclear structure functions $F^{WI}_{iA}(x,Q^2);~ (i=1-3)$ in the region of $x \le 1$ for the different values of $Q^2$ and Saito et al.~\cite{Saito:1985ct} have studied the effect of Fermi motion on 
$F^{EM}_{2A}(x,Q^2)$ for $x\ge1$ at higher values of $Q^2$ viz.
$Q^2=100$ GeV$^2$ by using the various forms for the momentum distribution of nucleons such as ideal Fermi gas type, 2-range Gaussian type, etc. in the different nuclear targets. 
Frankfurt et al.~\cite{Frankfurt:1988nt} have studied medium effects using the few nucleon correlation model as well as by considering the effect of short range correlations of nucleon on $F^{EM}_{2A}(x,Q^2)$
beyond $x=1$ and at moderate and high values of $Q^2$. Furthermore, Fernandez de Cordoba et al.~\cite{FernandezdeCordoba:1995pt} have also evaluated
the electromagnetic nuclear structure function $F^{EM}_{2A}(x,Q^2)$ for 
$x\gtrsim 1$ in the local density approximation by considering the effect of nucleon correlations. They obtained the results for carbon, oxygen and iron in the wide range of $Q^2$
and discussed the importance of DIS to understand the nucleon dynamics in the nuclear medium. One may notice that theoretical investigation of nuclear structure function in the 
region of low and moderate $Q^2$ at $x\gtrsim 1$ is lacking particularly for the experiments being performed using the (anti)neutrino beam which 
motivated us to carry out this study. In the present work, we have theoretically studied the nuclear medium effects in the DIS region for both the electromagnetic and weak interaction channels in the kinematic range of 
$x \gtrsim 1$, where the effect of Fermi motion and nucleon correlations come into play. These effects have been taken into account through the nucleon spectral function which provides
information about energy and momentum distribution of nucleons inside a nucleus, in a microscopic field theoretical model~\cite{Marco:1995vb, Frankfurt:1985ui}. To calculate the spectral function for an interacting Fermi sea in the nuclear medium 
we have used the nuclear many body theory and then the local density approximation (LDA) is applied to obtain the results for a finite nucleus. 
In LDA, nucleon density is calculated at the point of interaction for a volume element $d^3r$ inside the nuclear target and the free lepton-nucleon cross section is folded over the
density of the nucleons in the nucleus and integrated over the whole volume of the nucleus. This model has been applied earlier to understand the nuclear medium effects in both the electromagnetic 
and weak interaction channels up to $x\le 0.8$~\cite{SajjadAthar:2007bz, SajjadAthar:2009cr, Haider:2011qs, Haider:2012nf, Haider:2012ic, Haider:2015vea, Haider:2016tev, Haider:2016zrk, Zaidi:2019asc, Zaidi:2019mfd, Athar:2020kqn, Zaidi:2021iam}, where besides the nucleon-nucleon correlations and Fermi motion some other 
nuclear medium effects such as shadowing, antishadowing and mesonic cloud contributions are important. 

In this work, 
the numerical results are obtained for the charged lepton and neutrino induced DIS off carbon, argon, iron and lead nuclear targets by incorporating nuclear medium effects like binding energy, Fermi motion
and nucleon correlations along with the TMC effect and the PDFs evolution is done at NNLO. In the next section, we present the formalism for $l-A; (l=e^\pm/\mu^\pm,\nu_e/\nu_\mu)$ DIS in brief. In Sec.~\ref{res}, 
results are presented and discussed which is followed by the summary of this work in Sec.~\ref{sum}.
\section{Formalism}\label{formalism}
The general expression of the differential scattering cross section for lepton-nucleus deep inelastic scattering process
\begin{equation}\label{reac}
l^-(k)/\nu_l(k)+A(p_A) \rightarrow l^-(k^\prime) +X(p^\prime_A); ~~l=e~\textrm{or}~\mu,
\end{equation}
is given by~\cite{Zaidi:2019asc, Zaidi:2019mfd}
\begin{eqnarray}\label{d2sigdxdy_weak}
\frac { d^2\sigma_A^{IC} }{ dx dy }&=& \kappa \left[x y^2 F_{1A}^{IC}(x,Q^2) + \left(1-y-\frac{M_N x y}{ 2 E_l} \right) F_{2A}^{IC}(x,Q^2)  + x y \left(1-\frac{y}{2} \right)F_{3A}^{IC}(x,Q^2)\right]\;,
\end{eqnarray}
where in Eq.~\ref{reac}, $k(E_l,{\bf k})$ and $k'(E'_l,{\bf k'})$ are the four momenta of the incoming and outgoing leptons while $p_A(M_A,{\bf 0})$ and $p'_A(E_A',{\bf p_A'})$ are the four momenta of
the target nucleus and the final state jet of hadrons, respectively.
$M_A(=AM_N)$ is the mass of the target nucleus. In Eq.~\ref{d2sigdxdy_weak}, the superscript ``$IC$'' 
stands for the interaction channel which could be either the weak (WI) or electromagnetic (EM) interaction channel with $F_{3A}^{EM}(x,Q^2)=0$. The constant
$\kappa=\frac{8 M_N E_l \pi \alpha^2}{Q^4}$ for the EM interaction and $\kappa=\frac{G_F^2 M_N E_l}{\pi} \left( \frac{M_W^2}{M_W^2+Q^2}\right)^2$ for the weak interaction
induced processes, $\alpha$ is the strong coupling constant, $G_F$ is Fermi coupling constant,
$M_W$ is the mass of $W$ boson and $Q^2~(\ge 0)$ is the four momentum transfer square. 
$F_{iA}^{IC}(x,Q^2);~(i=1-3)$ are the dimensionless nuclear structure functions. 
The parity violating nuclear structure function $F_{3A}^{IC}(x,Q^2)$ arises due to the vector$-$axial vector interference part of 
the weak interaction and it does not contribute in the case of electromagnetic interaction. To evaluate the nuclear structure functions 
we perform the numerical calculations in the laboratory frame, where target nucleus is at rest ($p^0_A=M_A$, ${\bf p_A}=0$).
However, the nucleons bound inside the nucleus are not stationary but are moving with a momentum  ($p_N\ne 0$) constrained by the Fermi momentum ($p_{F_N}$) of the nucleon in the nucleus which is given by $p_N \le p_{F_N}$.
  In the global Fermi gas model, Fermi momentum  of nucleon is taken to be a 
constant value like $p_{F_N}=221$ MeV for carbon, $p_{F_N}=251$ MeV for iron, etc., while in the local density approximation, where the interaction takes place at a point ${\bf r}$
lying inside a volume element $d^3r$, instead of taking a constant density for a given nucleus the lepton scatters from a bound nucleon having density as
a function of $r$, i.e. $\rho_N(r)$ and the corresponding Fermi momentum is given by $p_{F_N}=(3 \pi^2 \rho_N(r))^{1/3}$. The differential scattering cross section which is evaluated as a function of local
density $\rho_N(r)$ is given by
\begin{equation}
 d\sigma_A=\int d^3r\;\rho_N(r)\;d\sigma_N, 
\end{equation}
where $d\sigma_N$ is the differential cross section of the lepton-nucleon scattering. For a symmetric nuclear matter, each nucleon occupies a volume of $(2\pi\hbar)^3$ and each unit cell is occupied by the 
two nucleons due to the two possible spin orientations. Hence the number of nucleons in a given volume $V$ are
\begin{equation}
 N=2 V\int_0^{p_{F_N}} \frac{d^3p}{(2\pi\hbar)^3} 
\end{equation}
In the natural unit system $\hbar=1$
\begin{equation}
 N=2 V\int_0^{p_{F_N}} \frac{d^3p}{(2\pi)^3}; \hspace{5mm}\textrm{or} \hspace{5mm} \rho=\frac{N}{V}=2 \int_0^{p_{F_N}} \frac{d^3p}{(2\pi)^3} \;n(\bf{p,r}),
\end{equation}
where $n(\bf{p,r})$ is the occupation number of a nucleon lying within the Fermi sea with the following constrains
\begin{equation}
 n(\bf{p,r})=\left\{
 \begin{array}{c}
 1~~;~~ p\le p_{F_N}\\
 0~~;~~  p>p_{F_N}
 \end{array}
 \right.
\end{equation}

Moreover, for a nonsymmetric nucleus such as argon, iron, lead, etc., we have taken into account the different densities for the 
proton ($\rho_p(r)$) and the neutron ($\rho_n(r)$) which are expressed as
\begin{equation}
 \rho_n(r)=\frac{A-Z}{A}\rho(r);~~~ \rho_p(r)=\frac{Z}{A}\rho(r),
\end{equation}
where $\rho(r)$ is the charged nuclear density and the corresponding Fermi momenta are given by
\begin{equation}
 p_{F_n}=(3 \pi^2 \rho_n(r))^{1/3};~~~ p_{F_p}=(3 \pi^2 \rho_p(r))^{1/3}
\end{equation}

For the nuclear charge density $\rho(r)$ different parameterizations are available in the literature such as harmonic oscillator density, modified harmonic oscillator density,
two-parameter Fermi density, three-parameter Fermi density, etc.~\cite{Vries, GarciaRecio:1991wk}. For the present numerical calculations, we have used modified harmonic oscillator (MHO) density for carbon while two-parameter
Fermi (2pF) density for argon, iron and lead which are given by
 \begin{eqnarray}
  \textrm{MHO ~density}:~~~\rho_N(r)&=&\rho_0\Big[1+c_2\left(\frac{r}{c_1} \right)^2 \Big]~,~~~\nonumber
 \textrm{2pF ~density}:~~~ \rho_N(r)=\frac{\rho_0}{1+e^{(r-c_1)/c_2}}\nonumber
 \end{eqnarray}
with $c_1$ and $c_2$ as the density parameters and $\rho_0$ as the central density~\cite{Vries, GarciaRecio:1991wk}. These parameters are individually tabulated in Table~\ref{table1}
for proton and neutron in the case of nonisoscalar nuclear target as well as for nucleon in the case of isoscalar nuclear target.
  \begin{table}
 \centering
\begin{tabular}{||c|cc|cc|c|c|c|c||} \hline\hline
 \multirow{2}{*}{ Nucleus}  & \multicolumn{4}{c|}{Nonisoscalar }   & \multicolumn{2}{c|}{Isoscalar} &$ \multirow{2}{*}{B.E./A}$ &$ \multirow{2}{*}{T/A}$\\\cline{2-7}
   &    $c_1^n$ & $c_1^p$ & $c_2^n$ & $c_2^p$  &  $c_1$   &  $c_2$& & \\\hline  
$^{12}$C  &  -  & -    &   - &    -   &  1.692& 1.082$^{\ast}$& 7.6   &  20.0 \\
$^{40}$Ar  &  3.64  & 3.47    &   0.569 &    0.569   &  3.53& 0.542& 8.6   &  29.0 \\
$^{56}$Fe &  4.050  & 3.971   &  0.5935    &   0.5935     & 4.106&0.519&  8.8    &  30.0   \\
$^{208}$Pb &  6.890   & 6.624    &  0.549   &  0.549     & 6.624&0.549&  7.8    & 32.6 \\ \hline \hline 
\end{tabular}
\caption{Different parameters used for the numerical calculations for various nuclei. For $^{12}$C
we have used modified harmonic oscillator density($^{\ast}$ $c_2$ is dimensionless) and for $^{40}$Ar, $^{56}$Fe and $^{208}$Pb nuclei,
 2-parameter Fermi density have been used, where superscript $n$ and $p$ in density parameters($c_{i}^{n,p}$; $i$=1,2) stand for neutron and proton, respectively. Density parameters for isoscalar and 
 nonisoscalar nuclear targets are given separately in units of fm.
 The kinetic energy per nucleon($T/A$) and the binding energy per nucleon ($B.E/A$) obtained using Eq.\ref{benergy}
 for the different nuclei are given in MeV. }
 \label{table1}
\end{table}
We have chosen the momentum transfer along the $z-$axis in the numerical calculations, i.e., $q^\mu=(q^0,0,0,q^z)$ leading to
 $x_N=\frac{Q^2}{2p_N \cdot q}=\frac{Q^2}{2 (p^0 q^0-p^z q^z)}$.
These bound nucleons also interact among themselves via the strong interaction. These effects have been taken into account for an inclusive process by using the 
hole spectral function ($S_h$) calculated in a microscopic field theoretical model~\cite{Marco:1995vb}. Moreover, we have ensured that the spectral function is properly normalized and
checked it by obtaining the correct baryon number ($A$) for a given nuclear target~\cite{Haider:2015vea}:
\begin{equation}\label{speca}
2 \int d^3 r\int \frac{d^3 p}{(2\pi)^3}\int_{-\infty}^{\mu}d\omega \; S_h(\omega,{\bf p})=A, 
\end{equation}
where the factor of 2 is a spin factor, $\omega(=p^0-M_N)$ is the removal energy and $\mu$ is the chemical potential defined in terms of Fermi momentum and the nucleon self energy ($\Sigma^N$) as~\cite{FernandezdeCordoba:1991wf}:
\begin{equation}\label{chem}
 \mu=\frac{p_{F_N}^2}{2M_N}+Re\Sigma^N\Big[\frac{p_{F_N}^2}{2M_N},p_{F_N} \Big]
\end{equation}
 The binding energy per nucleon for a nucleus~\cite{Haider:2015vea} is given by
\begin{equation}\label{benergy}
 |E_A|=-\frac{1}{2}\;\Big(<E_N-M_N>+\frac{A-2}{A-1}\;<T> \Big)
\end{equation}
with $<T>$ as the average kinetic energy and $<E_N>$ as total nucleon energy. 
Details are given in Ref.~\cite{Haider:2015vea, FernandezdeCordoba:1991wf}.

The spectral function has been calculated using the Lehmann's representation for the relativistic nucleon propagator and nuclear many body theory is used for calculating it for an interacting Fermi
sea in nuclear matter. Then the local density approximation is applied to translate these results to finite nuclei. 
 The details are given in Ref.~\cite{Haider:2015vea, Zaidi:2019asc}, where we have discussed that for an inclusive scattering process only the hole spectral 
 function ($S_h$) is required and the nuclear hadronic tensor ($W^{\mu \nu}_{A}$) is expressed in terms of the nucleon hole spectral function and the nucleonic
 hadronic tensor ($W^{\mu \nu}$) for an isoscalar nuclear target as
 \begin{equation}\label{conv_WAai}
W^{\mu \nu}_{A} = 4 \int \, d^3 r \, \int \frac{d^3 p}{(2 \pi)^3} \, 
\frac{M_N}{E_N ({\bf p})} \, \int^{\mu}_{- \infty} d p^0 S_h (p^0, {\bf p}, \rho(r))
W^{\mu \nu} (p, q), \,
\end{equation} 
where the factor of 4 is for spin-isospin of the nucleon. However, for a nonisoscalar nuclear target $W^{\mu \nu}_{A}$ is written in terms of the proton/neutron hole spectral function ($S_h^j;~j=p,n$) and the corresponding
hadronic tensor ($W^{\mu \nu}_{j};~j=p,n$) as
\begin{equation}\label{conv_WAa}
W^{\mu \nu}_{A} = 2\sum_{j=p,n} \int \, d^3 r \, \int \frac{d^3 p}{(2 \pi)^3} \, 
\frac{M_N}{E_N ({\bf p})} \, \int^{\mu_j}_{- \infty} d p^0 S_h^{j} (p^0, {\bf p}, \rho(r))
W^{\mu \nu}_{j} (p, q), \,
\end{equation}
where the factor of 2 is due to the two possible projections of nucleon spin and $\mu_j;~(j=p,n)$ is the chemical potential for the proton/neutron.
In LDA, the proton ($S_h^p$) and neutron ($S_h^n$) hole spectral functions are normalized separately to the respective proton and neutron numbers in a nuclear target
as~\cite{Haider:2015vea, Zaidi:2019asc}:
\begin{eqnarray}
\label{specp}
  2 \int d^3r\;\int \frac{d^3 p}{(2\pi)^3} \;\int_{-\infty}^{\mu_p}\;S_h^p(\omega,{\bf p},\rho_p(r))\;d\omega &=& Z\;, \\
  \label{specn}
    2 \int d^3r\;\int \frac{d^3 p}{(2\pi)^3} \;\int_{-\infty}^{\mu_n}\;S_h^n(\omega,{\bf p},\rho_n(r))\;d\omega &=& N\;.
 \end{eqnarray}
The hadronic tensor ($W^{\mu \nu}_{j}$) is written in terms of the dimensionless proton and neutron structure functions ($F_{ij}(x,Q^2);~i=1-3;~j=p,n$), therefore, by using Eq.~\ref{conv_WAa} and the general form of 
hadronic tensor with an appropriate choice of $x,y,z$ components, we obtain the following expressions of dimensionless nuclear structure functions 
for a nonisoscalar nuclear target~\cite{Haider:2015vea, Zaidi:2019mfd, Zaidi:2019asc}:
\begin{small}
\begin{eqnarray}
\label{conv_WA2weak}
F_{1A}^{IC}(x_A, Q^2) &=& 2\sum_{j=p,n} AM_N \int  d^3 r  \int \frac{d^3 p}{(2 \pi)^3} 
\frac{M_N}{E_N ({\bf p})}  \int^{\mu_j}_{- \infty} d p^0 ~S_h^j (p^0, {\bf p}, \rho_j(r))~\left[\frac{F_{1j}^{IC}(x_N, Q^2)}{M_N} + \left(\frac{p^x}{M_N}\right)^2 
\frac{F_{2j}^{IC}(x_N, Q^2)}{\nu_N}\right],\\
\label{had_ten151weak}
F_{2A}^{IC}(x_A,Q^2)  &=&  2\sum_{j=p,n} \int \, d^3 r \, \int \frac{d^3 p}{(2 \pi)^3} \, 
\frac{M_N}{E_N ({\bf p})} \, \int^{\mu_j}_{- \infty} d p^0 ~S_h^j (p^0, {\bf p}, \rho_j(r)) \times\left(\frac{M_N}{p^0~-~p^z~\gamma}\right) \times F_{2j}^{IC}(x_N,Q^2)\nonumber\\
&\times&\left[\left(\frac{Q}{q^z}\right)^2\left( \frac{|{\bf p}|^2~-~(p^{z})^2}{2M_N^2}\right) +  \frac{(p^0~-~p^z~\gamma)^2}{M_N^2} \left(\frac{p^z~Q^2}{(p^0~-~p^z~\gamma) q^0 q^z}~+~1\right)^2\right], \\
\label{f3aweak}
 F_{3A}^{IC}(x_A,Q^2) &=& 2 A \sum_{j=p,n} \int \, d^3 r \, \int \frac{d^3 p}{(2 \pi)^3} \, 
\frac{M_N}{E_N ({\bf p})} \, \int^{\mu_j}_{- \infty} d p^0 S_h^j (p^0, {\bf p}, \rho_j(r))\times \frac{q^0}{q^z} \left({p^0 q^z - p^z q^0  \over p \cdot q} \right)F_{3j}^{IC}(x_N,Q^2),
\end{eqnarray}
\end{small}
 where $\nu_N=\frac{p\cdot q}{M_N}=\frac{p^0 q^0 - p^z q^z}{M_N}$, $\gamma=\frac{q^0}{q^z}$. For an isoscalar nuclear target ($F_{iA}=\frac{F_{iA}^{p}+F_{iA}^{n}}{2}$) the factor of 2 in the 
 above expressions (Eqs.~\ref{conv_WA2weak}-\ref{f3aweak}) is replaced by 4 and $\mu_j$ is replaced by $\mu$ (see Eqs.~\ref{chem} and \ref{conv_WAai}).
 
 In the limit of $Q^2\to\infty$, $\nu\to \infty$ with $x\to$finite, the nucleon structure functions become the function of dimensionless Bjorken variable $x$ only, i.e., $F_{ij}^{IC}(x_N); ~(j=p,n)$ and 
 are expressed in terms of the parton distribution functions (PDFs) at the leading order (LO) as
 \begin{eqnarray}
  F_{2 j}^{IC}(x)&=& \sum_{i} \; e_i^2\;x\{q_i(x) +\bar q_i(x) \};~~~~
  x F_{3 j}^{IC}(x)= \sum_{i} \; e_i^2\;x\{q_i(x) -\bar q_i(x) \},
 \end{eqnarray}
where index $i$ runs over the flavor of quarks, $e_i$ is the charge of corresponding quark or antiquark and $x q_i(x)/ x \bar q_i(x)$ is the probability of finding a quark/antiquark inside
the nucleon carrying a momentum fraction $x$ of the momentum of the target nucleon. For the free nucleon case, Callan-Gross relation has been used to obtain $F_{1 j}^{IC}(x)$ in terms of PDFs at 
the leading order, i.e. $F_{2 j}^{IC}(x)=2 x F_{1 j}^{IC}(x)$. For $Q^2\to\infty$, the strong coupling constant ($\alpha_s(Q^2)$) is small and the terms beyond the leading order are negligible, but
at the finite values of $Q^2$, the strong coupling constant is large and the higher order terms like the next-to-leading order (NLO), next-to-next-to-leading order (NNLO), etc. can not be ignored. 
Hence, the dimensionless nucleon structure functions are written as a perturbative series expansion of the
strong coupling constant~\cite{vanNeerven:1999ca, vanNeerven:2000uj} as
\begin{equation}
 F_{iN}(x,Q^2)=\Big(\frac{\alpha_s(Q^2)}{4\pi}\Big)^0 F_{iN}^{LO}(x,Q^2)+\Big(\frac{\alpha_s(Q^2)}{4\pi}\Big)^1 F_{iN}^{NLO}(x,Q^2)+\Big(\frac{\alpha_s(Q^2)}{4\pi}\Big)^2 F_{iN}^{NNLO}(x,Q^2)+...
\end{equation}
In the present work, we have performed the evolution of PDFs up to next-to-next-to-the leading order (NNLO) and use the MMHT nucleonic PDFs parameterization~\cite{Harland-Lang:2014zoa} for the numerical calculations. 
Furthermore, at low and moderate values of $Q^2$ the nonperturbative effect of target mass corrections (TMC) comes into play which is important at high $x$. We have incorporated the TMC effect
following the operator product expansion approach~\cite{Schienbein:2007gr}. The target mass corrected nucleon structure functions are given by~\cite{Schienbein:2007gr, Zaidi:2019asc}
\begin{eqnarray}
 F_{1N}^{TMC}(x,Q^2)&=&F_{1N}(\xi)\left(\frac{x}{\xi \lambda}\right)(1+2\varrho(1-\xi)^2),\nonumber\\
 F_{2N}^{TMC}(x,Q^2)&=&F_{2N}(\xi)\left(\frac{x^2}{\xi^2 \lambda^3}\right)(1+6\varrho(1-\xi)^2),\nonumber\\
  F_{3N}^{TMC}(x,Q^2)&=&F_{3N}(\xi)\left(\frac{x}{\xi \lambda^2}\right)(1-\varrho(1-\xi) ln\xi),\nonumber
\end{eqnarray}
where $\varrho=\frac{M_N^2 x \xi}{Q^2 \lambda}$, $\lambda=\sqrt{1+\frac{4M_N^2 x^2}{Q^2}}$ and the Nachtmann variable $\xi=\frac{2 x}{1+\lambda}$.
 Following the present formalism, we have obtained the results of the nuclear structure functions which are required to evaluate 
 the results of the differential scattering cross sections. These numerical calculations are performed in the kinematic region of high $x(\gtrsim 0.8)$ and 
 moderate $Q^2 (\le 20$ GeV$^2$) as depicted in Fig.~\ref{xq2_fig} and the results are presented in Figs.~\ref{res1}-\ref{res5}.

\section{Results and Discussion}\label{res}
In this section, we present the results of the electromagnetic and weak nuclear structure functions using Eqs.~\ref{conv_WA2weak}-\ref{f3aweak}
as well as the differential cross 
sections using Eq.~\ref{d2sigdxdy_weak} relevant to the kinematic region of the charged lepton-nucleus scattering experiments such as CLAS, NE18, etc. and neutrino-nucleus scattering 
experiments like MINERvA and DUNE. All the 
numerical results are obtained for the deep inelastic scattering by incorporating the nuclear medium effects like the binding energy, Fermi motion and nucleon correlations through 
the use of hole spectral function. 
Theoretical results are obtained for the carbon, hydrocarbon, argon, iron and lead nuclear targets in the region 
of $x\gtrsim 0.8$ keeping $Q^2\ge 1$ GeV$^2$ without and with a cut on the center of mass energy $W$. Furthermore, in argon, iron and lead, isoscalarity corrections are also included wherever mentioned.
\begin{figure}
 \includegraphics[height=7 cm, width=7.5 cm]{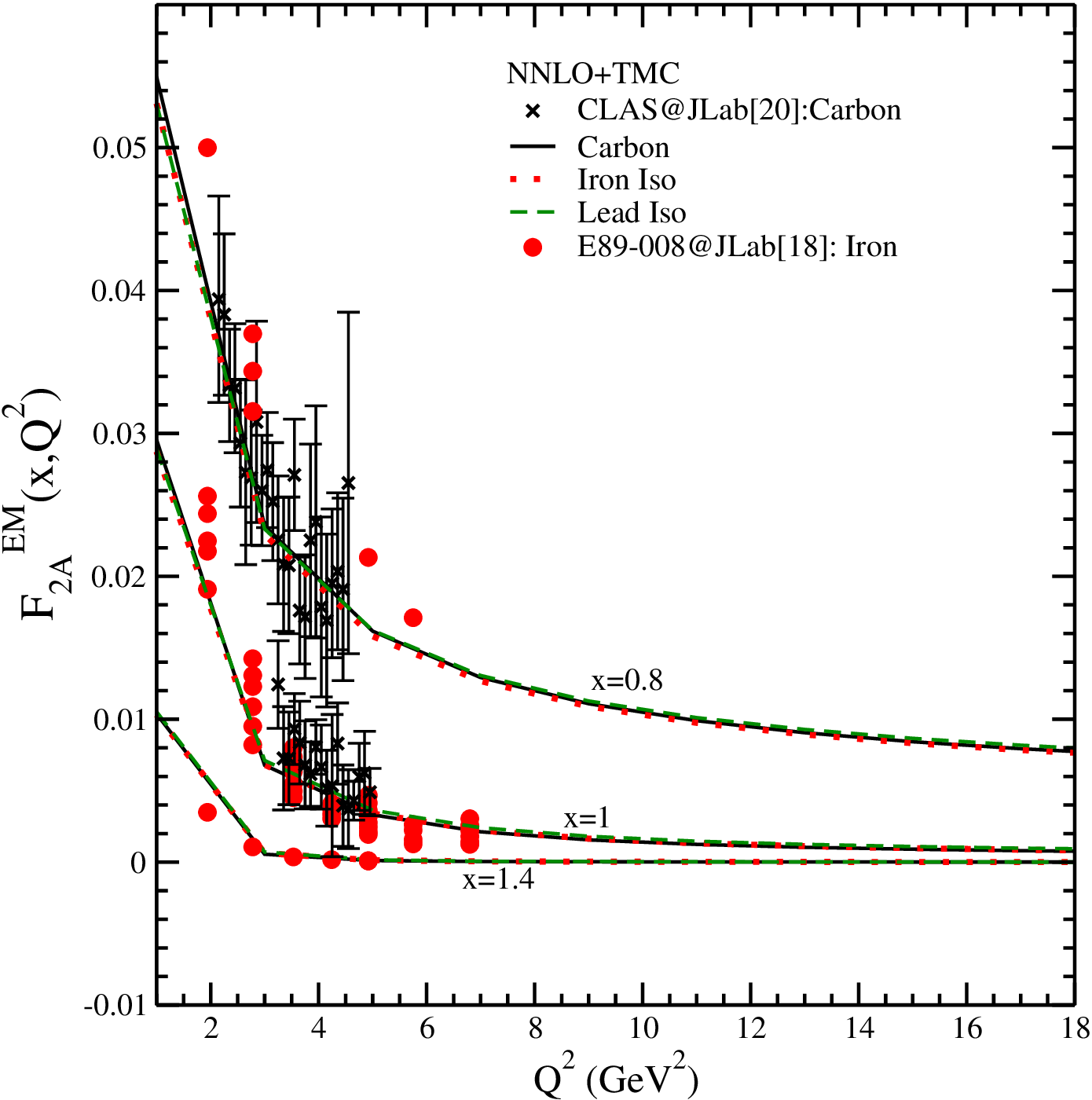}
  \includegraphics[height=7 cm, width=7.5 cm]{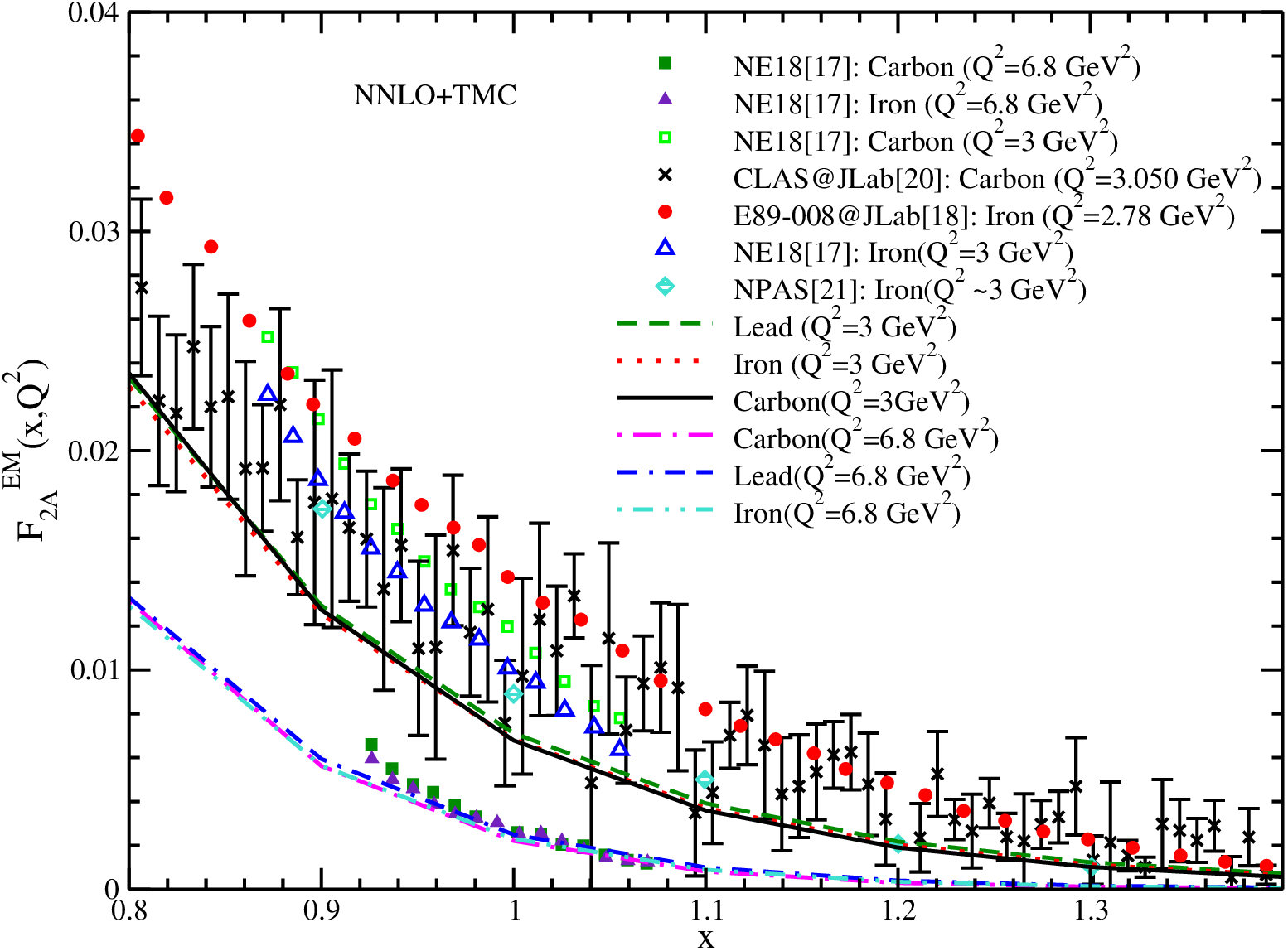}
 \caption{Results of electromagnetic nuclear structure function {\bf (Left panel:)} $F_{2A}^{EM}(x,Q^2)$ vs $Q^2$ at different values of $x$ and 
 {\bf (Right panel:)} $F_{2A}^{EM}(x,Q^2)$ vs $x$ at $Q^2=3$ and 6.8 GeV$^2$ are shown for carbon, iron and lead. These results are obtained
 at NNLO incorporating the TMC effect but without applying any cut on the center of mass energy $W$ and are compared with the experimental data of inclusive $e^--^{56}Fe$ scattering from 
NE18 experiment at NPAS~\cite{Arrington:1995hs}, E89-008 experiment at JLab~\cite{Arrington:2001ni}, experiment at NPAS~\cite{Filippone:1992iz} as well as 
with the experimental data for $e^--^{12}C$ from NE18 experiment at NPAS~\cite{Arrington:1995hs} and CLAS Hall B experiment at JLab~\cite{CLAS:2010nnk}. The theoretical curves are the results only for the DIS region while the experimental 
results are for the inclusive process.}
\label{res1}
\end{figure}

In the left panel of Fig.~\ref{res1}, the results of the electromagnetic nuclear structure function $F_{2A}^{EM}(x,Q^2)$ vs $Q^2$ for the nuclear targets $A=^{12}C,~^{56}Fe,~^{208}Pb$ 
are shown at the different values 
of $x$ viz. $x=0.8,~1.0$ and $1.4$. For the numerical calculations iron and lead are treated as isoscalar nuclear targets (Eq.~\ref{conv_WAai}). In this kinematic region the contribution from 
the deep inelastic channel to the total cross section of inclusive electron-nucleus scattering process is expected to be small as compared to the contributions from the inelastic resonance production and quasielastic scattering processes. Nevertheless,
the contribution from the deep inelastic region as may be observed from the figure is significant. With the increase in $x$, $F_{2A}^{EM}(x,Q^2)$ decreases and thus the contribution of DIS to the 
cross section becomes gradually small. The numerical results for $F_{2A}^{EM}(x,Q^2)$ are compared with the available experimental data for the inclusive
electron-nucleus scattering~\cite{Arrington:2001ni, CLAS:2010nnk} in the region of moderate $Q^2(\le 7$ GeV$^2$). 
In the right panel of the figure, the results are presented for $F_{2A}^{EM}(x,Q^2)$ vs $x$ at the two different values of $Q^2$ viz. $Q^2=3$ GeV$^2$ and $Q^2=6.8$ GeV$^2$ in carbon, iron and lead nuclear targets.
It may be observed that due to the $Q^2$ variation, there is significant difference in the results of nuclear structure function $F_{2A}^{EM}(x,Q^2)$, however, this difference
becomes small with the increase in $x$.
These results are also compared with the inclusive electron-nucleus scattering experimental data available for 
carbon~\cite{Arrington:1995hs, CLAS:2010nnk} and iron~\cite{Arrington:2001ni, Arrington:1995hs, Filippone:1992iz}.
One may notice that our theoretical results obtained using the present formalism for the DIS process underestimates the experimental data. It may be because of the missing contributions from the quasielastic and resonance production 
processes which we have not taken into account. Hence in order to understand the experimental results for the inclusive electron-nucleus scattering process, a theoretical study for the $Q^2$ dependence
of the inelastic resonance production and quasielastic processes should also be performed which is a topic of separate study.

\begin{figure}
 \includegraphics[height=7 cm, width=15 cm]{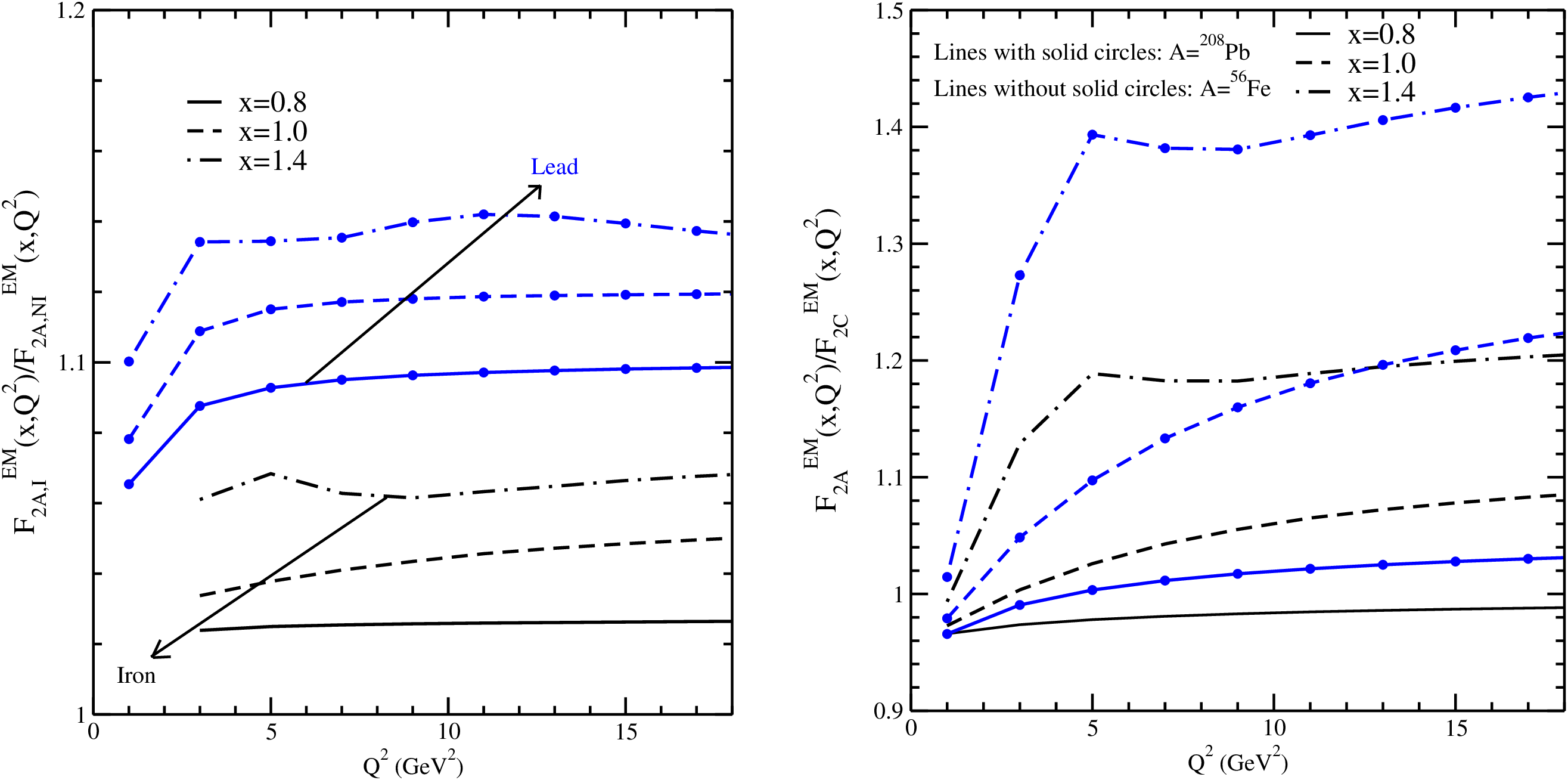}
 \caption{Results for the ratios of electromagnetic nuclear structure functions for isoscalar to nonisoscalar nuclear target 
 viz. $\frac{F_{2A,I}^{EM}(x,Q^2)}{F_{2A,NI}^{EM}(x,Q^2)};(I\equiv isoscalar,~NI\equiv nonisoscalar)$ (left panel) and 
 $\frac{F_{2A}^{EM}(x,Q^2)}{F_{2C}^{EM}(x,Q^2)}$ (isoscalar nuclear targets: right panel) are shown for iron and lead vs $Q^2$ at the different values of $x$. ``$A$'' represents the same nuclear target both 
 in the numerator and denominator. These results have been obtained using nucleonic PDFs at NNLO with TMC effect. }
\label{res1p}
\end{figure}
For the heavy nuclear targets like iron ($^{56}_{_{26}}Fe$, $N>Z$) and lead ($^{208}_{_{82}}Pb$, $N>>Z$) which have different neutron and proton numbers, isoscalarity 
corrections become important. 
Hence, it is required to observe the effect of the corrections arising due to neutron excess on nuclear structure functions 
for a given nuclear target by treating it to be isoscalar ($N=Z$) as well as nonisoscalar ($N\ne Z$).
In our theoretical model as discussed in Sec.~\ref{formalism} for a nonisoscalar nuclear target, the hole spectral function is separately normalized to the proton (Eq.~\ref{specp}) and
neutron (Eq.~\ref{specn}) numbers while for an isoscalar nuclear target $S_h$ is normalized to the nucleon numbers (Eq.~\ref{speca}).

To explicitly study the isoscalarity corrections we have obtained the results for the ratio $\frac{F_{2A,I}^{EM}(x,Q^2)}{F_{2A,NI}^{EM}(x,Q^2)};(I\equiv isoscalar,~NI\equiv nonisoscalar)$ vs $Q^2$ at 
a fixed value of $x$. These results are presented in the left panel of Fig.~\ref{res1p}. One may notice that the ratios viz. 
$\frac{F_{2Fe,I}^{EM}(x,Q^2)}{F_{2Fe,NI}^{EM}(x,Q^2)}$ and $\frac{F_{2Pb,I}^{EM}(x,Q^2)}{F_{2Pb,NI}^{EM}(x,Q^2)}$ have significant deviation from unity which highlights the 
importance of nonisoscalarity, especially in the heavier nuclear target like $^{208}Pb$. For example,
in lead ($N>>Z$) this nonisoscalarity effect is about $9-10\%$ in the entire range of 
$Q^2$ at $x=0.8$, while for $x=1.4$, this difference increases to $13-14\%$. Whereas for a nonisoscalar nuclear target such as iron, where $N\gtrsim Z$ this effect is small like $2-3\%$ at $x=0.8$ and 
$6-7\%$ at $x=1.4$. Except for low values of $Q^2\le 7$ GeV$^2$, this ratio is found to be almost $Q^2$ independent. Furthermore, to observe the nuclear medium modifications of $F_{2A}^{EM}(x,Q^2)$ in 
different nuclear targets the results for the ratios of iron to carbon $\Big(\frac{F_{2Fe}^{EM}(x,Q^2)}{F_{2C}^{EM}(x,Q^2)}\Big)$ and lead to carbon
$\Big(\frac{F_{2Pb}^{EM}(x,Q^2)}{F_{2C}^{EM}(x,Q^2)}\Big)$ have been obtained treating all the nuclear targets viz. $^{12}C$, $^{56}Fe$ and $^{208}Pb$ to be isoscalar. These results are shown in the right panel of the Fig.~\ref{res1p}. From the figure, one may observe that the
nuclear medium effects become more pronounced with the increase in mass number $A$, Bjorken $x$ as well as four momentum transfer square $Q^2$. Quantitatively, the increase in the nuclear medium effects in lead vs carbon is of about 
$\sim 5\%$, $16\%$ and $20\%$ at $Q^2=3$ GeV$^2$, $Q^2=9$ GeV$^2$ and $Q^2=15$ GeV$^2$, respectively when $x$ is kept fixed say here at $x=1.0$. With the increase in $x$, say at $x=1.4$ this difference
becomes $27\%$, $38\%$ and $42\%$ for the respective values of $Q^2$ as in the above case. In our earlier works on 
the study of nuclear medium effects in the DIS region~\cite{Haider:2015vea, Zaidi:2019asc, Zaidi:2019mfd} for $x$ lying in the range of $0\le x \le 0.8$, we found that the 
nuclear medium effect gradually decreases with the increase in $Q^2$. Present results show that the effects of the isoscalarity corrections and the nuclear medium modifications are significant even
at $x\gtrsim 1$.

\begin{figure}
 \includegraphics[height=6.5 cm, width=16 cm]{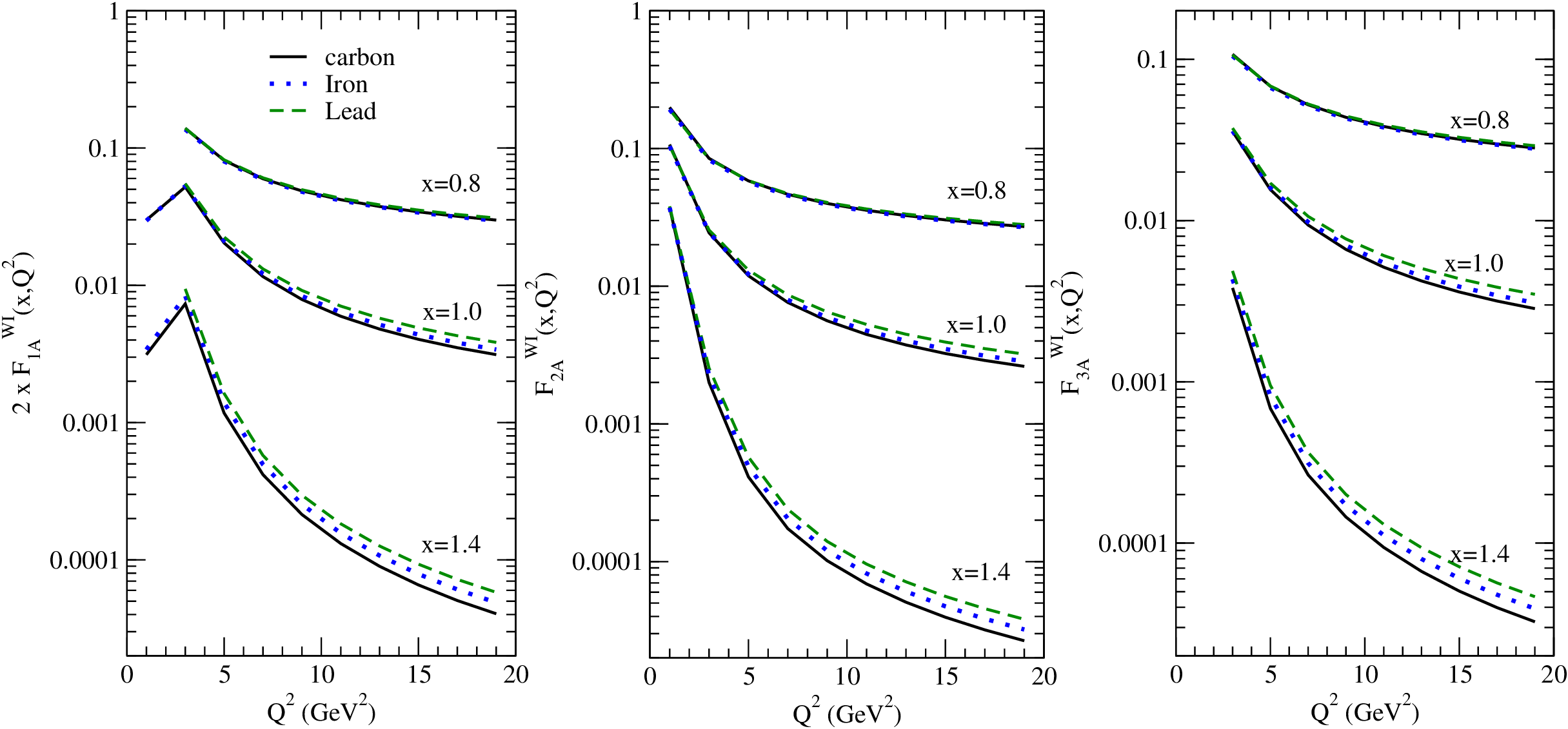}
 \caption{Results of weak nuclear structure functions $F_{iA}^{WI}(x,Q^2);~(i=1-3)$ vs $Q^2$ at the different values of $x$ are shown for carbon, iron and lead. These results are obtained
without applying any cut on the center of mass energy $W$ at NNLO incorporating the TMC effect. Iron and lead are treated as isoscalar nuclear targets. }
\label{res2}
\end{figure}

In Fig.~\ref{res2}, the numerical results for the weak nuclear structure functions $2 x F_{1A}^{WI}(x,Q^2)$, $F_{2A}^{WI}(x,Q^2)$ and $F_{3A}^{WI}(x,Q^2)$ are presented.  
It may be noticed that all the three nuclear structure functions decrease in magnitude with the increase in $x$ and $Q^2$.
We have also looked into the validity of Callan-Gross relation, i.e. $F_{2N}(x)=2 x F_{1N}(x)$, by comparing the results of $2 x F_{1A}^{WI}(x,Q^2)$ and $F_{2A}^{WI}(x,Q^2)$
in the presence of nuclear medium effects which have been discussed earlier in the case of electromagnetic nuclear structure functions in Ref.~\cite{Haider:2015vea, Zaidi:2019mfd} for $x<0.8$. CG relation holds good at the leading order for the free nucleon target, however, it
shows deviation at low and moderate values of $Q^2$ when gluonic contributions become significant beyond the leading order. In the present kinematic region of $x$ and $Q^2$, we find that there is significant deviation of the ratio $\frac{F_{2A}(x,Q^2)}{2 x F_{1A}(x,Q^2)}$ from 
unity (not shown here explicitly), for example, at $Q^2=3$ GeV$^2$ this deviation in the ratio $\frac{F_{2A}^{WI}(x,Q^2)}{2 x F_{1A}^{WI}(x,Q^2)}$ 
is $\sim 40\%$ at $x=0.8$, $52\%$ at $x=1.0$ and $72\%$ at $x=1.4$ which is almost independent of the nucleon mass number $A$. Whereas, at larger values of $Q^2$ viz. $Q^2=15$ GeV$^2$
this deviation decreases to $12\%$, $20\%$ and $40\%$ at $x=0.8$, $x=1.0$ and $x=1.4$, respectively.
The deviation of the ratio $\frac{F_{2A}(x,Q^2)}{2 x F_{1A}(x,Q^2)}$ from unity becomes more pronounced with the increase in $x$ and decrease in $Q^2$. 

\begin{figure}
 \includegraphics[height=6.5 cm, width=14 cm]{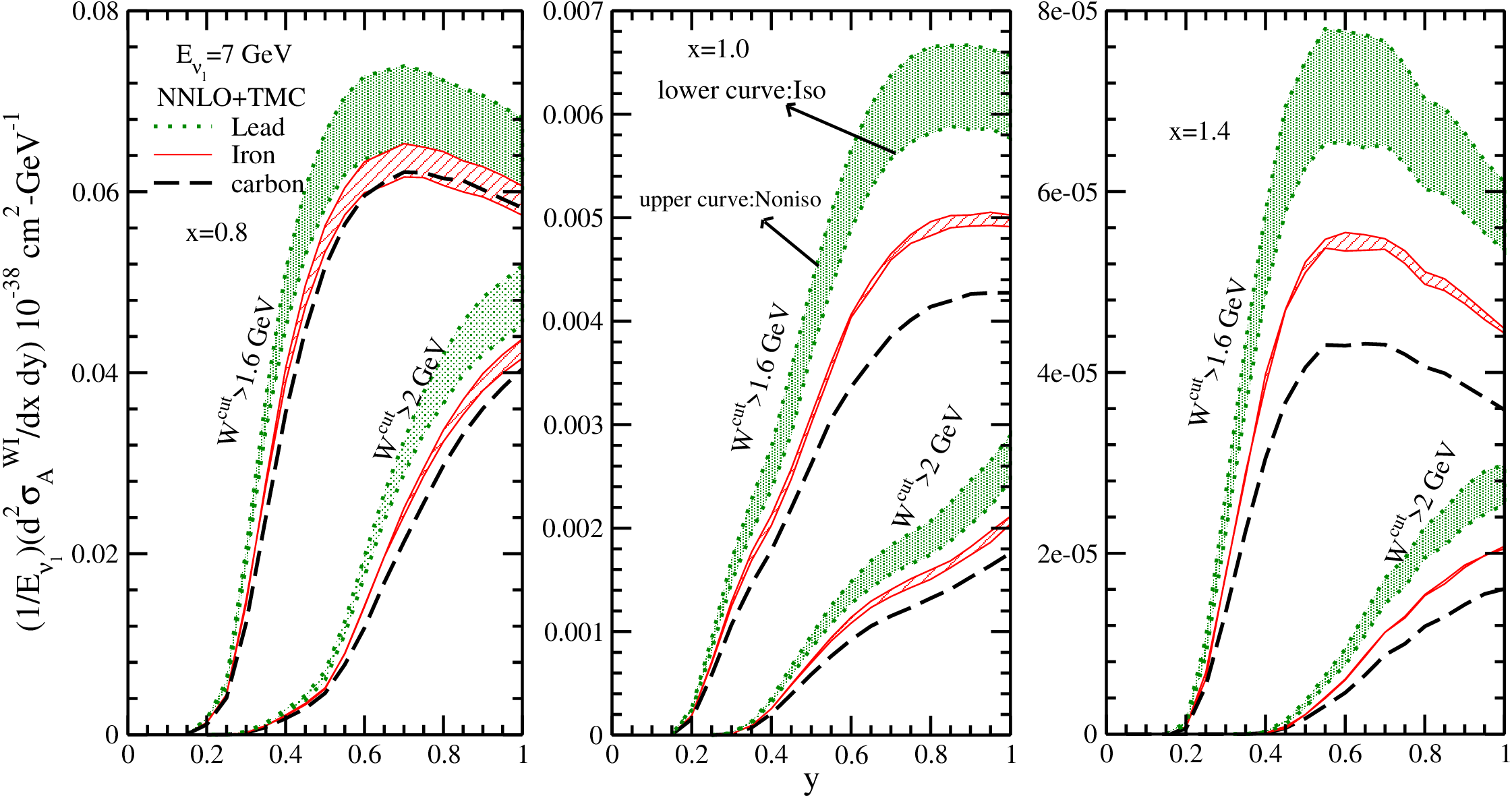}
 \caption{Results of $\frac{1}{E_{\nu_{l}}}\frac{d^2\sigma^{WI}_{A}}{dx dy}$ vs $y$ for $\nu_\mu-A;~(A=^{12}C,~^{56}Fe, ^{208}Pb)$ DIS are shown at NNLO with TMC effect. These results are obtained
 for $E_{\nu_{l}}=7$ GeV by applying the cuts of 1.6 GeV and 2.0 GeV on the CM energy, treating iron and lead isoscalar as well as nonisoscalar nuclear targets.}
 \label{res3}
\end{figure}

\begin{figure}
 \includegraphics[height=8 cm, width=14 cm]{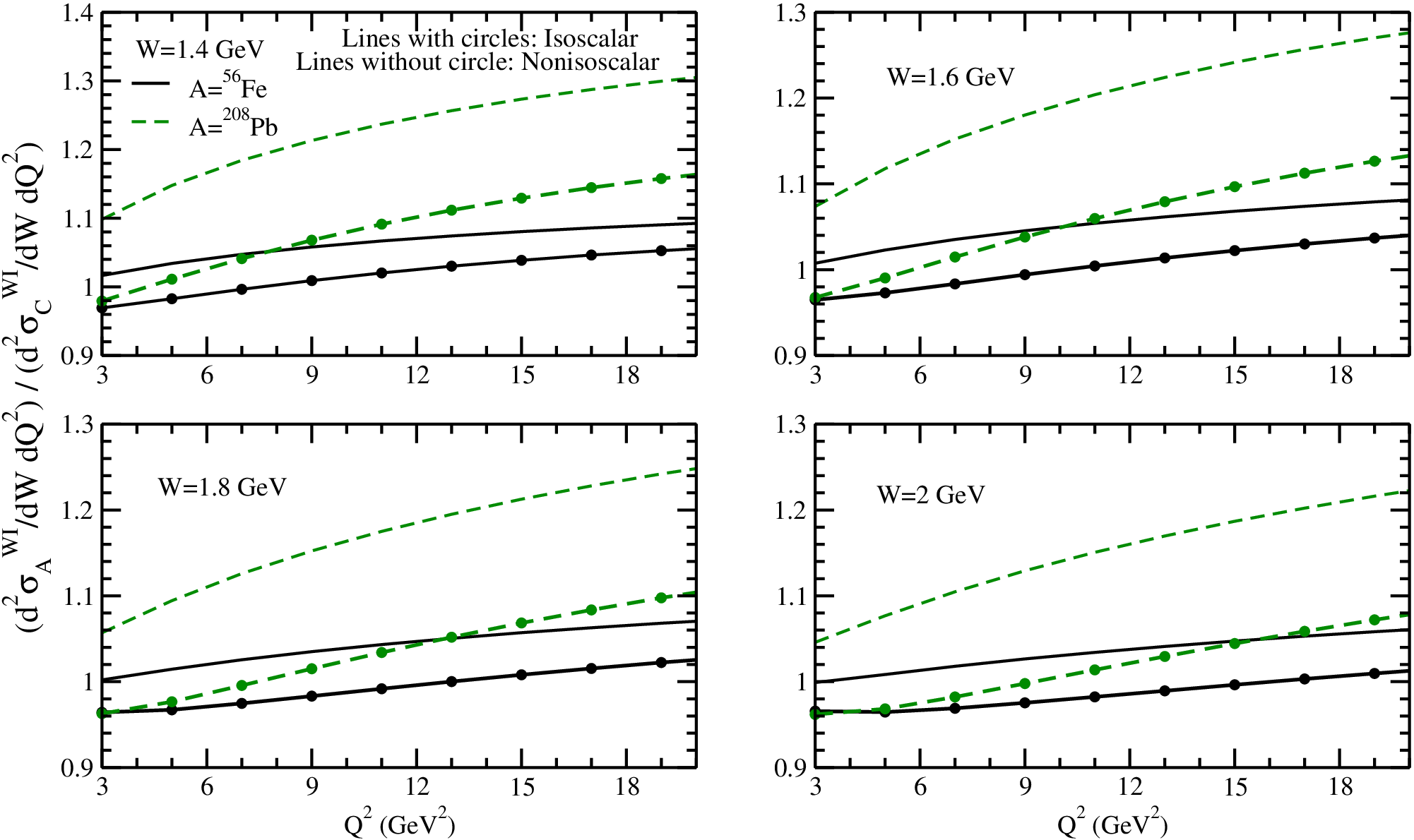}
  \caption{Results for the ratio of $\Big(\frac{d^2\sigma_A^{WI}}{dW dQ^2}\Big) / \Big(\frac{d^2\sigma_C^{WI}}{dW dQ^2}\Big);~~(A=^{56}Fe,~^{208}Pb)$ vs $Q^2$ for $\nu_\mu-A$ DIS are shown
  at NNLO with TMC effect.
  Nuclear targets iron and lead are treated to be isoscalar as well as nonisoscalar. Lines without solid circles represent the results for the nonisoscalar nuclei
  while lines with solid circles represent the results obtained for the isoscalar nuclear targets. Solid lines correspond to iron and dashed lines correspond to lead.}
 \label{res4}
\end{figure}

Using the results of weak nuclear structure functions $F_{iA}^{WI}(x,Q^2);i=1-3$, we have obtained the results 
for $\nu_\mu-A$ double differential scattering cross 
sections $\frac{1}{E_{\nu_{l}}}\frac{d^2\sigma_A^{WI}}{dx dy};~~(A=^{12}C,~^{56}Fe, ^{208}Pb)$ vs $y$ (using Eq.~\ref{d2sigdxdy_weak}), and the 
ratio of differential cross section for iron to carbon and lead to carbon, i.e., $\Big(\frac{d^2\sigma_A}{dW dQ^2}\Big) / \Big(\frac{d^2\sigma_C}{dW dQ^2}\Big);~~(A=^{56}Fe,~^{208}Pb)$ vs $Q^2$ by 
treating iron and lead both as an isoscalar as well as nonisoscalar nuclear targets. These results are shown 
in Figs.~\ref{res3} and \ref{res4}, respectively. In these results the effect of the center of mass energy cut on the differential scattering cross sections is also discussed.
The significance of CM energy cut in the region of $x\le 0.8$ have already been discussed by us in Refs.~\cite{Zaidi:2019asc, Zaidi:2019mfd, Zaidi:2021iam, SajjadAthar:2020nvy, Athar:2020kqn} and by the
other theoretical groups~\cite{Graczyk:2009px, Lalakulich:2006yn, Hagiwara:2003di, Kretzer:2002fr, Gazizov:2016dhn}, but for $x\gtrsim 1$ no study is available in the weak sector.

In Fig.~\ref{res3}, the numerical results of $\frac{1}{E_{\nu_{l}}}\frac{d^2\sigma_A^{WI}}{dx dy};~~(A=^{12}C,~^{56}Fe, ^{208}Pb)$ vs $y$, are presented at $E_{\nu_{l}}=7$ GeV with the 
kinematical cuts of $W^{cut}>1.6$ GeV (upper ones) and $W^{cut}>2$ GeV (lower ones). The numerical results for carbon are shown by the long dashed lines while the results for iron and lead 
are shown by the bands filled with the diagonal line pattern and the shaded pattern, respectively. The upper curve in the band is the results 
when a given nuclear target is treated to be nonisoscalar while the lower curve of the band is the result when nuclear target is treated as isoscalar. Hence, by using these bands one may easily 
quantify the nonisoscalarity effect. We have found that the enhancement in the results due to the nonisoscalarity effect is $14-15\%$ at $x=0.8$ and $y=0.4-0.6$ in lead and 
it increases to $18-19\%$ at $x=1.4$. It may be observed that the isoscalarity effect is not independent of $x$ and we find that it is also $Q^2$ dependent. 
We have also observed the effect of $W^{cut}$ on the differential scattering cross section by comparing the results corresponding to $W^{cut}>1.6$ GeV and $W^{cut}>2$ GeV. We have found that
when a cut of $W^{cut}>2$ GeV is applied, the results of the differential cross section for $\nu_\mu-^{56}Fe$ get reduced by about $75\%$ at $x=0.8-1.0$ and $y=0.6$ as compared to the results
obtained with $W^{cut}>1.6$ GeV. However, for $y=0.8$ this reduction becomes $46\%$ at $x=0.8$ and $68\%$ at $x=1.0$. It shows that the effect of $W^{cut}$ is quite significant in the present kinematic region of $x$.

Since it is important to understand both the effects of CM energy $W$ as well as $Q^2$ on the differential scattering cross sections, we have also performed the numerical calculations 
to observe the effect of $Q^2$ variation when $W$ is kept fixed. These results would be important to explicitly investigate the behavior of differential cross section in different regions of 
CM energy cut corresponding to the second resonance region up to the region of deep inelastic scattering.

In Fig.~\ref{res4}, the results for the ratios of the double differential scattering cross section, $\Big(\frac{d^2\sigma_A^{WI}}{dW dQ^2}\Big) / \Big(\frac{d^2\sigma_C^{WI}}{dW dQ^2}\Big);~~(A=^{56}Fe,~^{208}Pb)$ vs $Q^2$
are presented at fixed values of the center of mass energy. These results are obtained keeping the ongoing experimental analysis of MINERvA collaboration in mind
for $\nu_\mu/\bar\nu_\mu-$nucleus scattering in the region of $x\gtrsim1$~\cite{DanR}. It may be noticed from the figure that the nuclear medium effects become more pronounced for the 
heavier nuclear targets at all values of $W$. For example, the increase 
in the nuclear medium effects in the ratio  
$\Big(\frac{d^2\sigma_{Fe}^{WI}}{dW dQ^2}\Big) / \Big(\frac{d^2\sigma_C^{WI}}{dW dQ^2}\Big)$
obtained for $W$ = 1.4 GeV is about $3\%$ at $Q^2=3$ GeV$^2$ and $\sim6\%$ at $Q^2 = 20$ GeV$^2$. Whereas, in lead this ratio 
$\Big(\frac{d^2\sigma_{Pb}^{WI}}{dW dQ^2}\Big) / \Big(\frac{d^2\sigma_C^{WI}}{dW dQ^2}\Big)$ increase to about $17\%$ at $Q^2 = 20$ GeV$^2$. 
However, when the numerical calculations are performed at $W=2$ GeV, the effect of nuclear medium modifications becomes small, for example, it is found to be $2\%$ in
$\Big(\frac{d^2\sigma_{Fe}^{WI}}{dW dQ^2}\Big) / \Big(\frac{d^2\sigma_C^{WI}}{dW dQ^2}\Big)$ and  
$8\%$ in $\Big(\frac{d^2\sigma_{Pb}^{WI}}{dW dQ^2}\Big) / \Big(\frac{d^2\sigma_C^{WI}}{dW dQ^2}\Big)$ at $Q^2=20$ GeV$^2$. It shows that at higher values of $Q^2$ the nuclear medium modifications have 
a weak dependence on $A$. Furthermore, we have obtained these ratios 
by treating the nuclear targets to be nonisoscalar and the results are shown by the lines without solid circles. By comparing the results for the isoscalar vs nonisoscalar nuclear
targets, it may be observed that the nonisoscalarity effect is significant in the entire region of $Q^2$, however, the nonisoscalarity effect in these ratios of cross section decreases
with the increase in CM energy cut, for example, at $Q^2=3$ GeV$^2$ there is an enhancement (from isoscalarity) of about $\sim 12\%$ at $W=1.4$ GeV and $8\%$ at $W=2$ GeV. While at $Q^2=20$ GeV$^2$ this enhancement in the ratio is found to be $14\%$ irrespective of CM energy $W$. It shows that in the region of high $Q^2$
the nonisoscalarity effect becomes almost independent of $W$ considered in this work.

\begin{figure}
 \includegraphics[height=6 cm, width=12 cm]{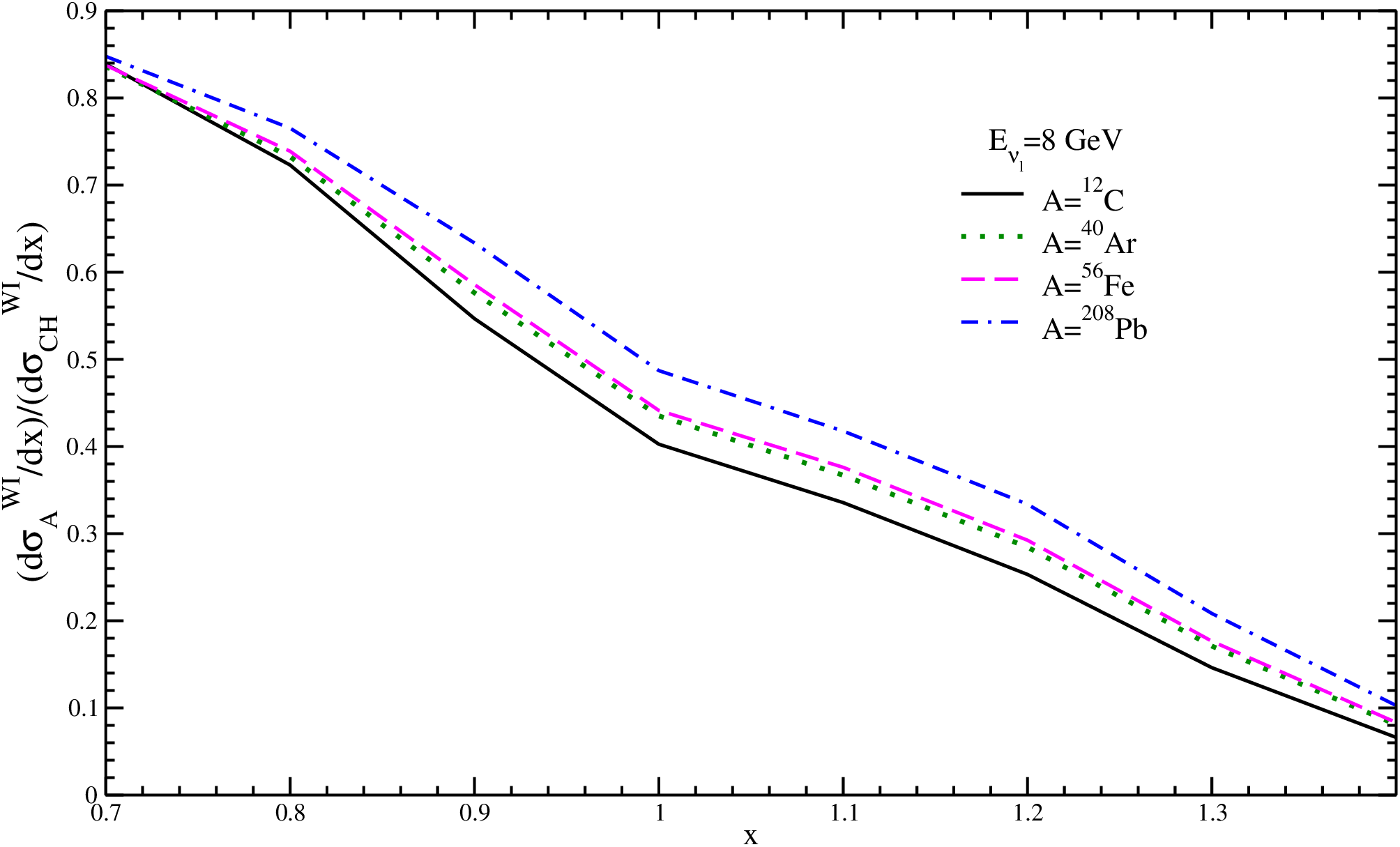}
  \caption{Results for the ratio of $\Big(\frac{d\sigma_A^{WI}}{dx}\Big) / \Big(\frac{d\sigma_{CH}^{WI}}{dx}\Big);~~(A=^{12}C,~^{40}Ar,~^{56}Fe,~^{208}Pb)$ vs $x$ for $\nu_\mu-A$ DIS 
  are shown at $E_{\nu_{l}}=8$ GeV keeping $Q^2>1$ GeV$^2$ and $W^{cut}>2$ GeV at NNLO with TMC effect.
  Nuclear targets argon, iron and lead are treated to be isoscalar.}
 \label{res5}
\end{figure}
In the short baseline neutrino experiments such as ICARUS~\cite{Tortorici:2019mwg, Machado:2019oxb} and SBND~\cite{MicroBooNE:2015bmn, SBND:2020scp} as well as in the 
long baseline neutrino experiment DUNE~\cite{Abi:2018dnh, Abi:2020mwi, Abi:2020qib},
liquid argon ($^{40}Ar$) is being used as nuclear target for the cross section measurements in the energy region of GeVs. However, the fixed target neutrino-nucleus scattering experiment MINERvA~\cite{MINERvA:2014rdw, Mousseau:2016snl}
is using hydrocarbon ($CH$), water ($H_2 O$), iron ($^{56}Fe$) and lead ($^{208}Pb$) nuclear targets and have recently reported the results showing the $x$-dependence of 
the inclusive cross section ratios i.e. $\frac{d\sigma_A/dx}{d\sigma_{CH}/dx}$ vs $x$ in the neutrino energy range of $2\le E_{\nu_{l}} \le 20$ GeV~\cite{MINERvA:2014rdw}. 
In Fig.~\ref{res5}, we have
explicitly shown the $x-$dependence of the 
nuclear medium effects for the deep inelastic scattering cross section ratios $\Big(\frac{d\sigma_A^{WI}}{dx}\Big) / \Big(\frac{d\sigma_{CH}^{WI}}{dx}\Big);~~(A=^{12}C,~^{40}Ar,~^{56}Fe,~^{208}Pb)$
at the neutrino energy $E_{\nu_{l}}=8$ GeV. These results are obtained with the kinematic
constrains of $Q^2>1$ GeV$^2$ and $W^{cut}>2$ GeV treating argon, iron and lead to be isoscalar nuclear targets. It is important to point out that the ratio of lead to hydrocarbon 
$\Big(\frac{d\sigma_{Pb}/dx}{d\sigma_{CH}/dx}\Big)$ is higher than the ratios $\frac{d\sigma_{Fe}/dx}{d\sigma_{CH}/dx}$, $\frac{d\sigma_{Ar}/dx}{d\sigma_{CH}/dx}$ and $\frac{d\sigma_{C}/dx}{d\sigma_{CH}/dx}$ as the Fermi motion effect is more 
pronounced in the heavier nuclear targets. For a meaningful comparison with the MINERvA's experimental results~\cite{MINERvA:2014rdw}, a separate study of $x-$dependence of the nuclear
medium effects in quasielastic and inelastic resonance 
production processes is also needed. This work is in progress and will be reported elsewhere. The theoretical predictions for argon would be relevant for the understanding of experimental results 
from DUNE~\cite{Abi:2020qib,Abi:2018dnh,Abi:2020mwi}.

\section{Summary and conclusions}\label{sum}
In this paper, the results of electromagnetic and weak nuclear structure functions have been presented along with the results of the differential scattering cross sections for the 
weak interaction induced $\nu_\mu-A$ deep inelastic scattering in the kinematic
region of high Bjorken $0.8\le x\le1.4$. These results are obtained at NNLO with the TMC effect, for carbon, hydrocarbon, argon, iron and lead. This study provides an overview of the nuclear 
medium modifications of the 
nucleon structure functions and the differential cross sections for the DIS process in the region of $x\gtrsim 1$ which has not been much explored yet.
Our findings are summarized as:
\begin{itemize}
 \item A comparison of the numerical results for $F_{2A}^{EM}(x,Q^2)$ with the 
inclusive electron-nucleus scattering experimental data for the electromagnetic nuclear structure function~\cite{Arrington:2001ni, Arrington:1995hs, Filippone:1992iz, CLAS:2010nnk} imply that for $x\gtrsim 1$, there is significant contribution from the
deep inelastic scattering region.
 \item Kinematic boundaries for the transition regions are needed to be precisely defined to distinguish between the resonance and DIS regions in order to avoid the double counting for 
 the measurements of the neutrino-nucleus scattering cross sections. We find that when a CM energy cut of $W^{cut}>2$ GeV vs $W^{cut}>1.6$ GeV is used to evaluate the $\nu_\mu-^{12}C$ 
 differential cross sections, then there is reduction in the DIS cross section which is more 
 pronounced in the region of low and intermediate $y$, like for $x=0.8$ this reduction is $80\%$ 
 at $y=0.6$ and $52\%$ at $y=0.8$. 
 The reduction in the cross section with $W^{cut}$ in the DIS region clearly shows that to determine the relative contributions of the 
inelastic resonance excitations and the DIS to the differential cross sections, especially in the region of low and intermediate $y$, proper kinematic constrains are required to be well understood.
 \item The effect of CM energy cut in the evaluation of cross section has also $x$ dependence, like at a fixed value of $y$ say, $y=0.6$, the differential cross section for $W^{cut}>2$ GeV vs $W^{cut}>1.6$ GeV gets reduced 
by $80\%$ at $x=0.8$ and $90\%$ at $x=1.4$ in $^{12}C$.
\item Isoscalarity corrections becomes more pronounced with the increase in $x$ and $(N-Z)/A$.
\item Theoretical predictions for $F_{iA}^{WI}(x,Q^2);~(i=1-3)$,
$\frac{d^2\sigma^{WI}_A}{dx dy}$, $\Big(\frac{d^2\sigma_A^{WI}}{dW dQ^2}\Big) / \Big(\frac{d^2\sigma_C^{WI}}{dW dQ^2}\Big)$ and $\Big(\frac{d\sigma_A^{WI}}{dx}\Big) / \Big(\frac{d\sigma_{CH}^{WI}}{dx}\Big)$
 presented in this work would be helpful in understanding the upcoming experimental results from the MINERvA collaboration and the planned DUNE experiment.
\end{itemize}

\section*{Acknowledgment}   
F. Zaidi is thankful to the Council of Scientific \& Industrial Research (CSIR), India, for providing the research associate fellowship with 
award letter no. 09/112(0622)2K19 EMR-I.
M. S. A. is thankful to the Department of Science and Technology (DST), Government of India for providing 
financial assistance under Grant No. SR/MF/PS-01/2016-AMU/G.

\end{document}